\documentclass[12pt,leqno]{article}

\title{Post-collapse dynamics of self-gravitating \\
Brownian particles and bacterial populations}

\usepackage{amsthm,amsmath,amssymb,a4wide,psfig,epsf}
\def\mb#1{\setbox0=\hbox{$#1$}\kern-.025em\copy0\kern-\wd0
\kern-0.05em\copy0\kern-\wd0\kern-.025em\raise.0233em\box0}

\begin{document}

\author{Cl\'ement Sire and Pierre-Henri Chavanis}
\maketitle
\begin{center}
Laboratoire de Physique Th\'eorique (FRE 2603 du CNRS), Universit\'e
Paul Sabatier,\\ 118, route de Narbonne, 31062 Toulouse Cedex 4, France\\
E-mail: {\it Clement.Sire{@}irsamc.ups-tlse.fr ~\&~
Chavanis{@}irsamc.ups-tlse.fr }

\vspace{0.5cm}
\end{center}

\begin{abstract}

We address the post-collapse dynamics of a self-gravitating gas of
Brownian particles in $D$ dimensions, in both canonical and
microcanonical ensembles. In the canonical ensemble, the post-collapse
evolution is marked by the formation of a Dirac peak with increasing
mass. The density profile outside the peak evolves self-similarly with
decreasing central density and increasing core radius. In the
microcanonical ensemble, the post-collapse regime is marked by the
formation of a ``binary''-like structure surrounded by an almost
uniform halo with high temperature. These results are consistent with
thermodynamical predictions in astrophysics. We also show that the
Smoluchowski-Poisson system describing the collapse of
self-gravitating Brownian particles in a strong friction limit is
isomorphic to a simplified version of the Keller-Segel equations
describing the chemotactic aggregation of bacterial
populations. Therefore, our study has direct applications in this
biological context.

\end{abstract}

\section{Introduction}
\label{sec_introduction}

Self-gravitating systems such as globular clusters and elliptical
galaxies constitute a Hamiltonian system of particles in interaction that
can be supposed isolated in a first approximation \cite{bt}. Since
energy is conserved, the proper description of stellar systems is the
microcanonical ensemble \cite{paddy}. The dynamical evolution of
elliptical galaxies is governed by the Vlasov-Poisson system which
corresponds to a collisionless regime. On the other hand, the kinetic
theory of globular clusters is based on the Landau-Poisson system (or
the orbit averaged Fokker-Planck equation) which describes a
collisional evolution. These equations conserve mass and energy.
Furthermore, the Landau equation increases the Boltzmann entropy
(H-theorem) due to stellar encounters. These equations have been
studied for a long time in the astrophysical literature and a
relatively good physical understanding has now been achieved
\cite{bt}. In particular, globular clusters can experience core
collapse \cite{larson,cohn,lbe,lance} related to the ``gravothermal
catastrophe'' concept \cite{lbw}.

For systems with long-range interactions, statistical ensembles are
not equivalent \cite{paddy}. Therefore, it is of conceptual interest
to compare the microcanonical evolution of stellar systems to a
canonical one in order to emphasize the analogies and the
differences. This can be achieved by considering a gas of
self-gravitating Brownian particles \cite{charosi} subject to a
friction originated from the presence of an inert gas and to a
stochastic force (modeling turbulent fluctuations,
collisions,...). This system has a rigorous canonical structure where
the temperature $T$ measures the strength of the stochastic force. Thus, we can precisely check the thermodynamical predictions  of Kiessling \cite{kiessling} and Chavanis \cite{chavcano} obtained in the canonical ensemble. In
the mean-field approximation, the self-gravitating Brownian gas model
is described by the Kramers-Poisson system. In a strong friction limit
(or for large times) it reduces to the Smoluchowski-Poisson
system. These equations conserve mass and decrease the Boltzmann free
energy \cite{gt}. They possess a rich physical and mathematical
structure and can lead to a situation of ``isothermal collapse'',
which is the canonical version of the gravothermal catastrophe.  These
equations have not been considered by astrophysicists because the
canonical ensemble is not the correct description of stellar systems
and usual astrophysical bodies do not experience a friction with a gas
(except dust particles in the solar nebula \cite{planete}).  Yet, it
is clear that the self-gravitating Brownian gas model is of
considerable conceptual interest in statistical mechanics to
understand the strange thermodynamics of systems with long-range
interactions and the inequivalence of statistical ensembles. In
addition, it provides one of the first model of stochastic particles
with long-range interactions, thereby extending the classical
Einstein-Smoluchowski model \cite{risken} to a more general context \cite{gt}.

In addition, it turns out that the same type of equations occur in
biology in relation with the chemotactic aggregation of bacterial
populations \cite{murray}. A general model of chemotactic aggregation
has been proposed by Keller \& Segel \cite{keller} in the form of two
coupled partial differential equations. In some approximation, this
model reduces to the Smoluchowski-Poisson system
\cite{charosi}. Therefore, there exists an {\it isomorphism} between
self-gravitating Brownian particles and bacterial colonies. Non-local
drift-diffusion equations analogous to the Smoluchowski-Poisson system
have also been introduced in two-dimensional hydrodynamics in relation
with the formation of large-scale vortices such as Jupiter's great red
spot \cite{rs,csr,houches}. These analogies give further physical
interest to our Brownian model.

In a recent series of papers \cite{charosi,cs1,cs2}, we have studied
the dynamics and thermodynamics of self-gravitating Brownian particles
confined within a spherical box of radius $R$ in a space of dimension
$D$.  In these works, we focused on the pre-collapse regime. In the
canonical situation (fixed temperature $T$) we showed that a critical
temperature $T_{c}$ exists below which the system undergoes a
gravitational collapse leading to a finite time singularity at
$t=t_{coll}$. For $t\rightarrow t_{coll}$, the evolution is
self-similar in the sense that the density profile evolves as
$\rho(r,t)=\rho_{0}(t)f(r/r_{0}(t))$ where $f(x)$ is independent on
time. For $x\rightarrow +\infty$, $f(x)\sim x^{-\alpha}$ with
$\alpha=2$. The central density increases as $\rho_{0}\sim
(t_{coll}-t)^{-1}$ and the core radius decreases as $r_{0}\sim
(t_{coll}-t)^{1/\alpha}$.  For $T=0$, the exponent is
$\alpha=2D/(D+2)$. The scaling profiles can be calculated
analytically. These results can be compared with those of Penston
\cite{penston} who considered an isothermal collapse modelled by the
Euler-Jeans equations. We also introduced a microcanonical description
of Brownian particles by letting the temperature $T(t)$ evolve in time
so as to conserve energy. This can provide a simplified model for the
violent relaxation of collisionless stellar systems
\cite{csr} or, simply, a numerical algorithm  \cite{gt} to determine what is the maximum entropy state at fixed $E$ and $M$. In the microcanonical 
situation, there exists a critical energy $E_{c}$ (Antonov energy)
below which the system collapses. For $t\rightarrow t_{coll}$, there
exists a pseudo-scaling regime where $\alpha$ passes very slowly from
$\alpha_{max}=2.21...$ to $\alpha=2$. Numerical simulations suggest
that $T(t)$ remains finite at $t=t_{coll}$ so that the true scaling
regime corresponds to $\alpha=2$, as in the canonical situation
\cite{cs2,guerra}.

What happens after $t_{coll}$~?  By investigating the case $T=0$, we
found in \cite{cs1} that the evolution continues in the post-collapse
regime with the formation of a Dirac peak accreting more and more mass
as $M(t)\sim (t-t_{coll})^{D/2}$ while the density outside the peak
evolves self-similarly with decreasing central density $\rho_{0}\sim
(t-t_{coll})^{-1}$ and increasing core radius $r_{0}\sim
(t-t_{coll})^{D+2\over 2D}$. Our aim in this paper is to investigate
the post-collapse regime for $T\neq 0$ in both canonical and
microcanonical ensembles. In Sec. \ref{sec_analogy}, we emphasize the
analogy between self-gravitating Brownian particles and bacterial
populations. In the following, we shall use the astrophysical
terminology but we stress that our results apply
equally well to biology where the chemotactic model has more concrete
physical applications. In Sec. \ref{sec_collapse}, we set the
notations and recall the main results concerning the pre-collapse
dynamics. In Sec. \ref{sec_postT0}, we study the post-collapse
dynamics at $T=0$ by a method different from \cite{cs1}, which can be
generalized at finite temperature. The post-collapse dynamics at $T>0$
is precisely considered in Sec. \ref{sec_postT}. In the canonical
ensemble, we show that the system forms a Dirac peak whose mass
increases as $M(t)\sim (t-t_{coll})^{D/2-1}$ while the density profile
for $r>0$ expands self-similarly with $\rho_{0}\sim (t-t_{coll})^{-1}$
and $r_{0}\sim (t-t_{coll})^{1/2}$. For large times, the system is
made of a Dirac peak of mass $\sim M$ surrounded by a light gas of
Brownian particles (with negligible self-interaction). Due to thermal
motion, complete collapse takes an infinite time (contrary to the case
$T=0$). For $t\rightarrow +\infty$, the mass contained in the Dirac
peak increases as $1-M(t)/M\sim {\rm exp}(-\lambda t)$ where $\lambda$
is the fundamental eigenvalue of a quantum problem. For $T\rightarrow
0$, we find that $\lambda=1/4T+c_{D}/T^{1/3}+...$.  In the
microcanonical ensemble, the post-collapse regime is very
pathological. The system tends to create a ``Dirac peak of $0^{+}$
mass'' surrounded by a uniform halo with infinite temperature. The
central structure is reminiscent of a ``binary star'' containing a
weak mass $2m\ll M=Nm$ but a huge binding energy comparable to the
potential energy of the whole cluster. Since our model is essentially
mean-field, the physical formation of binaries is replaced by the type
of structures mentioned above. The formation of a ``Dirac peak''
containing the whole mass in canonical ensemble and the formation of
``binaries'' in microcanonical ensemble are expected from
thermodynamical considerations
\cite{antonov,kiessling,paddy,chavcano,fermions,cs1}. 
These are the structures which provoke a divergence of free energy (at
fixed mass and temperature) and entropy (at fixed mass and energy)
respectively \cite{cs1}. All these analytical results are confirmed by
numerical simulations of the Smoluchowski-Poisson system in
Secs. \ref{sec_ns} and
\ref{sec_postmicro}. We were able in particular to ``cross the
singularity'' at $t=t_{coll}$ and describe the post-collapse dynamics.

\section{Analogy between self-gravitating Brownian particles and bacterial populations}
\label{sec_analogy}

\subsection{Self-gravitating Brownian particles}
\label{sec_sgb} 

We  consider a system of self-gravitating
Brownian particles described by the $N$ coupled stochastic equations
\begin{eqnarray}
{d{\bf r}_i\over dt}={\bf v}_i, \quad {d{\bf v}_i\over dt}=-\xi
{\bf v}_i-\nabla_i U({\bf r}_1,...,{\bf r}_N)+\sqrt{2D'}\ {\bf R}_i(t),
 \label{brown1bis}
\end{eqnarray}
where $\xi$ is the friction coefficient, $D'$ is the diffusion
coefficient and ${\bf R}_{i}(t)$ is a white noise satisfying $\langle
{\bf R}_{i}(t)\rangle={\bf 0}$ and $\langle
{R}_{a,i}(t){R}_{b,j}(t')\rangle=\delta_{ij}\delta_{ab}\delta(t-t')$,
where $a,b=1,...,D$ refer to the coordinates of space and
$i,j=1,...,N$ to the particles. The particles interact via the
potential $U({\bf r}_{1},...,{\bf r}_{N})=\sum_{i<j}u({\bf r}_{i}-{\bf
r}_{j})$. In this paper, $u({\bf r}_{i}-{\bf r}_{j})$ is the Newtonian
binary potential in $D$ dimensions. The stochastic process (\ref{brown1bis}) defines a {\it toy model} of
gravitational dynamics which extends the classical Brownian model
\cite{risken} to the case of stochastic particles in interaction. In
this context, the friction is due to the presence of an inert gas and
the stochastic force is due to classical Brownian motion, turbulence,
or any other stochastic effect.

Starting from the $N$-body Fokker-Planck equation and using a
mean-field approximation \cite{martzel,bedlewo}, we can derive the
nonlocal Kramers equation
\begin{eqnarray}
\label{brown2} {\partial f\over\partial t}+{\bf v}{\partial
f\over\partial {\bf r}}+{\bf F}{\partial f\over\partial {\bf v}}=
{\partial\over\partial {\bf v}}\biggl ( D'{\partial f\over\partial
{\bf v}}+\xi f {\bf v}\biggr ),
\end{eqnarray}
where ${\bf F}=-\nabla\Phi$ is the smooth gravitational force felt by
the particles. The  gravitational potential $\Phi$ is related to the density $\rho=\int f d^{D}{\bf v}$ by the Poisson equation
\begin{eqnarray}
\Delta\Phi=S_{D}G\rho,
 \label{po}
\end{eqnarray}
where $S_D$ is the surface of the unit $D$-dimensional
sphere. Equation (\ref{brown2}) can be considered as a generalized
version of the Kramers-Chandrasekhar equation introduced in a
homogeneous medium \cite{chandra}. In this work, the diffusion and the
friction model stellar encounters in a simple stochastic
framework. The condition that the Maxwell-Boltzmann distribution is a
stationary solution of Eq. (\ref{brown2}) leads to the Einstein
relation $\xi=D'\beta$ where $\beta=1/T$ is the inverse temperature
(we have included the mass of the particles and the Boltzmann constant
in the definition of $T$). In our case, we do {\it not} assume that
the medium is homogeneous, so that we have to solve the
Kramers-Poisson system. This makes the study much more complicated
than usual. Up to date, we do not know any astrophysical application
of this model although there could be connexions with the process of
planetesimal formation in the solar nebula \cite{planete}. Whatever,
this model is interesting to develop on a conceptual point of view
because it possesses a rigorous thermodynamical structure and presents
the same features as more realistic models (isothermal distributions,
collapse, phase transitions,...). For this Brownian model, the
relevant ensemble is the canonical one since the temperature $T$ is
fixed.  Therefore, the Kramers-Poisson system can be viewed as the
canonical counterpart of the Landau-Poisson system. It is interesting
to study these two models in parallel to illustrate dynamically the
inequivalence of statistical ensembles for systems with long-range
interactions.

To simplify the problem further, we shall consider the strong
friction limit $\xi\rightarrow +\infty$, or equivalently the limit
of large times $t\gg \xi^{-1}$. In that approximation, we can
neglect the inertia of the particles. Then, the coupled stochastic
equations (\ref{brown1bis}) simplify to
\begin{eqnarray}
\xi {d{\bf r}_i\over dt}=-\nabla_i U({\bf r}_1,...,{\bf r}_N)+\sqrt{2D'}\ {\bf
R}_i(t).
 \label{brown3}
\end{eqnarray}
Furthermore, to leading order, the velocity distribution is
Maxwellian
\begin{eqnarray}
f({\bf r},{\bf v},t)=\biggl ({\beta\over 2\pi}\biggr )^{D/2}\rho({\bf r},t)e^{-\beta {v^{2}\over 2}},
 \label{brown4}
\end{eqnarray}
and the Kramers equation reduces to the Smoluchowski
equation
\begin{eqnarray}
\label{brown5} {\partial\rho\over\partial t}=\nabla \biggl \lbrack
{1\over\xi}(T\nabla\rho+\rho\nabla\Phi)\biggr\rbrack.
\end{eqnarray}
It can be shown that the Kramers equation decreases the Boltzmann
free energy 
\begin{eqnarray}
\label{brown6}  F[f]=E-TS=\int f {v^{2}\over 2}d^{D}{\bf r}d^{D}{\bf v}+{1\over 2}\int\rho\Phi d^{D}{\bf r}+T\int f\ln f d^{D}{\bf r}d^{D}{\bf v},
\end{eqnarray}
i.e. $\dot F\le 0$ and $\dot F=0$
at statistical equilibrium (canonical $H$-theorem). Similarly, the
Smoluchowski equation decreases the free energy $F[\rho]$ which is
obtained from $F[f]$ by using the fact that the velocity
distribution is Maxwellian in the strong friction limit. This
leads to the classical expression
\begin{eqnarray}
\label{brown7} F[\rho]=T\int \rho\ln\rho \ d^{D}{\bf r}+{1\over 2}\int
\rho\Phi \ d^{D}{\bf r}.
\end{eqnarray}
The passage from the Kramers equation to the Smoluchowski equation in
the strong friction limit is classical \cite{kampen}. It can also be
obtained formally from a Chapman-Enskog expansion \cite{cll}. 

In order to prevent finite time singularities and infinite densities,
we can consider a model of self-gravitating Brownian fermions,
enforcing the constraint $f\le \eta_{0}$ (Pauli exclusion
principle). The corresponding Kramers equation takes the form
\begin{eqnarray}
\label{brown2bis} {\partial f\over\partial t}+{\bf v}{\partial
f\over\partial {\bf r}}+{\bf F}{\partial f\over\partial {\bf v}}=
{\partial\over\partial {\bf v}}\biggl \lbrack D'{\partial f\over\partial
{\bf v}}+\xi f(\eta_{0}-f) {\bf v}\biggr \rbrack.
\end{eqnarray} 
In the strong friction limit, we obtain a Smoluchowski
equation of the form
\begin{eqnarray}
\label{brown8} {\partial\rho\over\partial t}=\nabla \biggl \lbrack
{1\over\xi}(\nabla p+\rho\nabla\Phi)\biggr\rbrack,
\end{eqnarray}
where $p(\rho)$ is the equation of state of the Fermi gas. The fermionic 
Smoluchowski-Poisson system has been studied in \cite{ribot}. Generalized Kramers and Smoluchowski equations are introduced in \cite{gt}.

\subsection{The Keller-Segel model} \label{sec_conc}

The name chemotaxis refers to
the motion of organisms (amoeba) induced by chemical signals
(acrasin). In some cases, the biological organisms secrete a
substance that has an attractive effect on the organisms
themselves. Therefore, in addition to their diffusive motion, they
move systematically along the gradient of concentration of the
chemical they secrete (chemotactic flux). When attraction prevails
over diffusion, the chemotaxis can trigger a self-accelerating
process until a point at which aggregation takes place. This is
the case for the slime mold {\it Dictyostelium Discoideum} and for
the bacteria {\it Escherichia coli} \cite{murray}.

A model of slime mold aggregation has been introduced by Keller \&
Segel \cite{keller} in the form of two coupled differential equations
\begin{equation}
\label{ge1} {\partial\rho\over\partial t}=\nabla(D_{2}\nabla\rho)-\nabla (D_{1}\nabla c),
\end{equation}
\begin{equation}
\label{ge2}{\partial c\over\partial t}=-k(c)c+f(c)\rho+D_{c}\Delta c.
\end{equation}
In these equations $\rho({\bf r},t)$ is the concentration of
amoebae and $c({\bf r},t)$ is the concentration of acrasin.
Acrasin is produced by the amoebae at a rate $f(c)$. It can also
be degraded at a rate $k(c)$. Acrasin diffuse according to Fick's
law with a diffusion coefficient $D_c$. Amoebae concentration
changes as a result of an oriented chemotactic motion in the
direction of a positive gradient of acrasin and a random motion
analogous to diffusion. In Eq. (\ref{ge1}), $D_2(\rho,c)$ is the diffusion
coefficient of the amoebae and $D_1(\rho,c)$ is a measure of the
strength of the influence of the acrasin gradient on the flow of
amoebae. This chemotactic drift is the fundamental process in the
problem.

A first simplification of the Keller-Segel model is provided by
the system of equations
\begin{equation}
\label{ge3} {\partial\rho\over\partial t}=D\Delta\rho-\chi\nabla (\rho\nabla c),
\end{equation}
\begin{equation}
\label{ge4}{\partial c\over\partial t}=D'\Delta c+a \rho-b c,
\end{equation}
where the parameters are positive constants. An additional
simplification, introduced by J\"ager \& Lauckhaus \cite{jager}
consists in ignoring the time derivative in Eq. (\ref{ge4}). This is valid in
the case where the diffusion coefficient $D'$ is large. Taking also
$b=0$, we obtain
\begin{equation}
\label{ge5} {\partial\rho\over\partial t}=D\Delta\rho-\chi\nabla (\rho\nabla c),
\end{equation}
\begin{equation}
\label{ge6}\Delta c=-\lambda\rho,
\end{equation}
where $\lambda=a/D'$. Clearly, these equations are isomorphic to the
Smoluchowski-Poisson system (\ref{brown5})-(\ref{po}) describing
self-gravitating Brownian particles in a strong friction limit. In
particular, the chemotactic flux plays the same role as the
gravitational drift in the overdamped limit of the Brownian
model. When chemotactic attraction prevails over diffusion, the system
is unstable and the bacteria start to aggregate. This blow-up is
similar to the collapse of self-gravitating systems in a canonical
situation. We note that in the Keller-Segel model, the diffusion
coefficient can depend on the density, leading to anomalous
diffusion. Such a situation is considered in
\cite{cs2} where  the nonlinear Smoluchowski-Poisson
system is studied.

The Keller-Segel model ignore clumping and sticking effects.  However,
at the late stages of the blow-up, when the density of amoebae has
reached high values, finite size effects and stickiness must clearly
be taken into account. As a first step, we can propose \cite{bedlewo}
to replace the classical equation (\ref{ge5}) by an equation of the form
\begin{equation}
\label{ge7} {\partial\rho\over\partial t}=D\Delta\rho-\chi\nabla (\rho(\sigma_{0}-\rho)\nabla c),
\end{equation}
which enforces a limitation $\rho({\bf r},t)<\sigma_{0}$ on the
maximum density of amoebae. This is the counterpart of the model of
self-gravitating Brownian fermions \cite{ribot}. These type of non-local
Fokker-Planck equations also occur in 2D hydrodynamics and
astrophysics in relation with the formation of large-scale vortices
and galaxies \cite{csr,houches}. Their systematic study is clearly of
broad interest \cite{gt}.

\section{Collapse dynamics of self-gravitating Brownian particles}
\label{sec_collapse}

\subsection{The Smoluchowski-Poisson system}
\label{sec_sp}

At a given temperature $T$ controlling the diffusion coefficient, the
density $\rho({\bf r},t)$ of self-gravitating Brownian particles satisfies the
following coupled equations:
\begin{eqnarray}
{\partial\rho\over\partial t}&=&\nabla\biggl \lbrack
{1\over\xi}(T\nabla\rho+\rho\nabla\Phi)\biggr\rbrack,
\label{brown1}\\
\Delta\Phi&=&S_{D}G\rho,
\end{eqnarray}
where $\Phi$ is the gravitational potential, and $S_D$ is the surface
of the unit $D$-dimensional sphere.

From now on, we set $M=R=G=\xi=1$ and we restrict ourselves to
spherically symmetric solutions. The equations of the problem become
\begin{eqnarray}
{\partial\rho\over\partial t}&=&\nabla (T\nabla\rho+\rho\nabla\Phi),
\label{dim1}\\
\Delta\Phi&=&S_{D}\rho,
\label{dim2}
\end{eqnarray}
with proper boundary conditions in order to impose a vanishing
particle flux on the surface of the confining sphere. These read
\begin{equation}
{\partial\Phi\over\partial r}(0,t)=0, \qquad \Phi(1)={1\over 2-D},
\qquad T{\partial \rho\over\partial r}(1)+\rho(1)=0,
\label{dim4}
\end{equation}
for $D>2$. For $D=2$, we take $\Phi(1)=0$ on the boundary. Integrating
Eq.~(\ref{dim2}) once, we can rewrite the Smoluchowski-Poisson system
in the form of a single integrodifferential equation
\begin{equation}
{\partial\rho\over\partial t}={1\over r^{D-1}}{\partial\over\partial
r}\biggl\lbrace r^{D-1}\biggl (T{\partial\rho\over\partial
r}+{\rho\over r^{D-1}}\int_{0}^{r}\rho(r')S_{D}r^{'D-1}dr'\biggr
)\biggr \rbrace.
\label{dim5a}
\end{equation}
The total energy is given as the sum of the kinetic and potential
contributions
\begin{equation}
E={D\over 2}T+{1\over 2}\int \rho\Phi d^{D}{\bf r}.
\label{dim3}
\end{equation}

The Smoluchowski-Poisson system is also equivalent to a single
differential equation
\begin{equation}
\frac{\partial M}{\partial t}=T \left(\frac{\partial^2 M}{\partial r^2}
-\frac{D-1}{r}\frac{\partial M}{\partial r}\right)
+{1\over r^{D-1}}M\frac{\partial M}{\partial r},
\label{sca}
\end{equation}
for the quantity
\begin{equation}
M(r,t)=\int_{0}^{r}\rho(r')S_{D}r^{'D-1}\,dr', \label{dint}
\end{equation}
which represents the mass contained within the sphere of radius
$r$. The appropriate boundary conditions are
\begin{equation}
 M(0,t)=N_0(t),\qquad M(1,t)=1,
\label{dintb}
\end{equation}
where $N_0(t)=0$, except if the density develops a condensed Dirac peak
contribution at $r=0$, of total mass $N_0(t)$.  It is also convenient to
introduce the function $s(r,t)=M(r,t)/r^{D}$ satisfying
\begin{equation}
{\partial s\over\partial t}=T\biggl ({\partial^{2}s\over\partial
r^{2}}+{D+1\over r}{\partial s\over\partial r}\biggr )+\biggl
(r{\partial s\over\partial r}+Ds\biggr )s.
\label{seq}
\end{equation}

\subsection{Self-similar solutions of the Smoluchowski-Poisson system}
\label{sec_sss}

In \cite{charosi,cs1,cs2}, we have shown that in the canonical
ensemble (fixed $T$), the system undergoes gravitational collapse
below a critical temperature $T_c$ depending on the dimension of
space.  The density develops a scaling profile, and the central
density grows and diverges at a finite time $t_{coll}$. The case $D=2$
was extensively studied in
\cite{cs1} and turns out to be very peculiar. Throughout this paper,
we restrain ourselves to the more generic case $D>2$, although other
dimensions play a special role as far as static properties are
concerned (see \cite{cs1,cs2}).

We look for self-similar solutions of the form
\begin{equation}
\rho(r,t)=\rho_{0}(t)f\biggl ({r\over r_{0}(t)}\biggr ), \qquad r_{0}=
\biggl ({T\over \rho_{0}}\biggr )^{1/2},
\label{dim5}
\end{equation}
where the King's radius $r_0$ defines the size of the dense core \cite{bt}.
In terms of the mass profile, we have
\begin{equation}
M(r,t)=M_{0}(t)g\biggl ({r\over r_{0}(t)}\biggr ), \qquad {\rm
with}\qquad M_{0}(t)=\rho_{0}r_{0}^{D},
\label{dim6}
\end{equation}
and
\begin{equation}
g(x)=S_{D}\int_{0}^{x}f(x')x^{'D-1}\,dx'. \label{dim7}
\end{equation}
In terms of the function $s$, we have
\begin{equation}
s(r,t)=\rho_{0}(t)S\biggl ({r\over r_{0}(t)}\biggr ), \qquad {\rm
with}\qquad S(x)={g(x)\over x^{D}}.
\label{dim6s}
\end{equation}

Substituting the {\it ansatz} (\ref{dim6s}) into Eq.~(\ref{seq}), we find that
\begin{equation}
{d\rho_{0}\over dt}S(x)-{\rho_{0}\over r_{0}}{dr_{0}\over dt}x
S'(x)={\rho_{0}^{2}}\biggl (S''(x)+{D+1\over
x}S'(x)+xS(x)S'(x)+DS(x)^{2}\biggr ),
\label{dim8}
\end{equation}
where we have set $x=r/r_{0}$. The variables of position and time
separate provided that $\rho_{0}^{-2}d\rho_{0}/dt$ is a constant
that we arbitrarily set equal to 2.  After time integration, this leads to
\begin{equation}
\rho_{0}(t)={1\over 2}(t_{coll}-t)^{-1},
\label{dim11}
\end{equation}
so that the central density becomes infinite in a finite time $t_{coll}$.
The scaling equation now reads
\begin{equation}
2S+xS'=S''+{D+1\over x}S'+S(xS'+DS).
\label{scalingd}
\end{equation}
The scaling solution of Eq.~(\ref{scalingd}) was obtained analytically 
in \cite{cs1} and reads
\begin{equation}
S(x)=\frac{4}{D-2+x^2},
\label{solscad}
\end{equation}
which decays with an exponent $\alpha=2$. This leads to
\begin{equation}
f(x)=\frac{4(D-2)}{S_D}\frac{x^2+D}{(D-2+x^2)^{2}} ,\qquad
g(x)=\frac{4x^D}{D-2+x^2}.
\end{equation}

Note finally that within the core radius $r_0$, the total mass  in
fact vanishes as $t\to t_{coll}$. Indeed, from  Eq.~(\ref{dim6}), we obtain
\begin{equation}
M(r_0(t),t)\sim \rho_{0}(t)r_{0}^{D}(t) \sim T^{D/2}(t_{coll}-t)^{D/2-1}.
\label{m0}
\end{equation}
Therefore, the collapse does {\it not} create a Dirac peak (``black hole'').

In \cite{cs1}, we have also studied the collapse dynamics at $T=0$
for which we obtained
\begin{equation}
\rho_0(t)\sim S_D^{-1}(t_{coll}-t)^{-1},
\end{equation}
as previously, but the core radius is not given anymore by the
King's radius which vanishes for $T=0$. Instead, we find
\begin{equation}
r_0\sim \rho_0^{-1/\alpha},
\end{equation}
with
\begin{equation}
\alpha=\frac{2D}{D+2}.
\end{equation}
The scaling function $S(x)$ is only known implicitly
\begin{equation}
\left\lbrack\frac{2}{D+2}-S(x)\right\rbrack^{\frac{D}{D+2}}=K x^{\frac{2D}{D+2}}S(x),
\label{st0}
\end{equation}
where $K$ is a known constant (see \cite{cs1} for details), $S(0)=\frac{2}{D+2}$, and the
large $x$ asymptotics $S(x)\sim f(x)\sim x^{-\alpha}$. The mass within the core radius is now
\begin{equation}
M(r_0(t),t)\sim \rho_{0}(t)r_{0}^{D}(t) \sim (t_{coll}-t)^{D/2},
\label{m0t0}
\end{equation}
and it again tends to zero as $t\rightarrow t_{coll}$.  Comparing
Eq.~(\ref{m0}) and Eq.~(\ref{m0t0}) suggests that if the temperature
is very small, an apparent scaling regime corresponding to the $T=0$
case will hold up to a cross-over time $t_*$, with
\begin{equation}
t_{coll}-t_*\sim T^{D/2}.\label{tst0}
\end{equation}
Above $t_*$, the $T\ne 0$ scaling ultimately prevails.

\section{Post-collapse dynamics at $T=0$}
\label{sec_postT0}

So far, all studies concerning the collapse dynamics of
self-gravitating Brownian particles have concentrated on the time
period $t\leq t_{coll}$. A natural question arises: what is happening
for $t>t_{coll}$~? The first possible scenario is that the state
reached at $t=t_{coll}$ is in fact a stationary state. However, its is
easy to check (see \cite{charosi}) that this is absolutely not the
case. In addition, the preceding study leads to a sort of paradox
\cite{fermions}. Indeed, we know that the statistical equilibrium state in the canonical ensemble is a Dirac peak \cite{kiessling,chavcano}. This is not the structure that forms at $t=t_{coll}$. This structure is singular at the origin ($\rho\sim r^{-2}$) but different from a Dirac peak (in particular the central mass is zero). This means that the evolution {\it must} continue after $t_{coll}$. In particular, we will show that the Dirac peak predicted by statistical mechanics forms in the post-collapse regime.

The scenario that we are now exploring is the following. A central
Dirac peak containing a mass $N_0(t)$ emerges at $t>t_{coll}$, whereas
the density for $r>0$ satisfies a scaling relation of the form
\begin{equation}
\rho(r,t)=\rho_{0}(t)f\biggl ({r\over r_{0}(t)}\biggr ),
\label{rhpost}
\end{equation}
where $\rho_0(t)$ is now decreasing with time (starting from
$\rho_0(t=t_{coll})\rightarrow +\infty$) and $r_0(t)$ grows with time
(starting from $r_0(t=t_{coll})=0$).  As time increases, the residual
mass for $r>0$ is progressively swallowed by the dense core made of
particles which have fallen on each other. It is the purpose of the
rest of this paper to show that this scenario actually holds, as well
as to obtain analytical and numerical results illustrating this final
collapse stage.

In this section, we present an alternative treatment to that of
\cite{cs1}, where this scenario was analytically shown to hold at
$T=0$. This new approach is a good introduction to the general $T\ne
0$ case which is studied in the next section.
We refer the reader to \cite{cs1} for an explicit solution of the $T=0$
post-collapse regime, which we found, leads to a central peak containing
all the mass in a finite time $t_{end}$.

For $T=0$, the dynamical equation for the integrated mass $M(r,t)$ reads
\begin{equation}
\frac{\partial M}{\partial t}=
{1\over r^{(D-1)}}M\frac{\partial M}{\partial r},
\label{scaT0}
\end{equation}
with  boundary conditions
\begin{equation}
 M(0,t)=N_0(t),\qquad M(1,t)=1.
\end{equation}
We define  $\rho_0$ such that for small $r$
\begin{equation}
 M(r,t)-N_0(t)=\rho_0(t)\frac{r^D}{D}+...
\end{equation}
Up to the geometrical factor $S_D^{-1}$, $\rho_0(t)$ is the
central residual density (the residual density is defined as the
density after the central peak has been subtracted). For $r=0$,
Eq.~(\ref{scaT0}) leads to the evolution equation for $N_0$
\begin{equation}
\frac{d N_0}{d t}=\rho_0 N_0.
\label{N0T0}
\end{equation}
As $N_0(t)=0$ for $t\leq t_{coll}$, and since this equation is a
first order differential equation, it looks like $N_0(t)$ should
remain zero for $t> t_{coll}$ as well. However, since
$\rho_0(t_{coll})=+\infty$, there is mathematically speaking no
global solution for this equation and non zero values for $N_0(t)$
can emerge from Eq.~(\ref{N0T0}), as will soon become clear.

We then define
\begin{equation}
s(r,t)=\frac{M(r,t)-N_0(t)}{r^D},
\end{equation}
which satisfies
\begin{equation}
{\partial s\over\partial t}=\biggl
(r{\partial s\over\partial r}+Ds\biggr )s+\frac{N_0}{r^D}\biggl
(r{\partial s\over\partial r}+Ds-\rho_0\biggr ).
\label{sT0}
\end{equation}
By definition, we have also $s(0,t)=\rho_0(t)/D$. 

We now look for a scaling solution of the form
\begin{equation}
s(r,t)=\rho_{0}(t)S\biggl ({r\over r_{0}(t)}\biggr ),
\label{st0post1}
\end{equation}
with $S(0)=D^{-1}$ and
\begin{equation}
\rho_0(t)= r_0(t)^{-\alpha},\label{r0t0}
\end{equation}
where $r_0$ is thus defined without ambiguity.  Inserting this scaling
{\it ansatz} in Eq. (\ref{sT0}), and defining the scaling variable
$x=r/r_0$, we find
\begin{equation}
{1\over \alpha\rho_0^{2}}\frac{d\rho_0}{dt}\left(\alpha S+x
S'\right)= S(DS+xS')+ \frac{N_0}{\rho_0r_0^D}{1\over x^{D}}(DS+xS'-1).
\label{scasT0}
\end{equation}
Imposing scaling, we find that both time dependent coefficients
appearing Eq.~(\ref{scasT0}) should be in fact constant. We thus
define a constant $\mu$ such that
\begin{equation}
N_0=\mu \rho_0r_0^D,
\end{equation}
and set
\begin{equation}
{1\over \alpha\rho_0^{2}}\frac{d\rho_0}{dt}=-\kappa, \label{kappa}
\end{equation}
with $\kappa>0$, as the central residual density is expected to
decrease.  Equation (\ref{kappa}) implies that $\rho_0\sim
(t-t_{coll})^{-1}$, which along with Eq.~(\ref{r0t0}) implies that
$N_0\sim (t-t_{coll})^{D/\alpha-1}$. We thus find a power law behavior
for $N_0$, which in order to be compatible with Eq.~(\ref{N0T0}), leads
to
\begin{equation}
\rho_0(t)=\left(\frac{D}{\alpha}-1\right)(t-t_{coll})^{-1},
\end{equation}
and then to
\begin{equation}
\kappa=\frac{1}{D-\alpha}.
\end{equation}
We end up with the scaling equation
\begin{equation}
\frac{1}{D-\alpha}\left(\alpha S+x S'\right)+ S(DS+xS')+\mu
x^{-D}(DS+xS'-1)=0. \label{scas1T0}
\end{equation}
From Eq.~(\ref{scas1T0}), we find that the large $x$ asymptotics
of $S$ is $S(x)\sim x^{-\alpha}$. In a short finite time after
$t_{coll}$, it is clear that the large distance behavior of the
density profile ($r\gg r_0$) cannot dramatically change. We deduce
that the decay of $S$ should match the behavior for time slightly
less than $t_{coll}$ for which $S(x)\sim x^{-\frac{2D}{D+2}}$.
Hence the value of $\alpha$ should remain unchanged before and
after $t_{coll}$. Finally, we obtain the following exact behaviors
for short time after $t_{coll}$:
\begin{eqnarray}
\rho_0(t)&=&\frac{D}{2}(t-t_{coll})^{-1},\\
r_0(t)&=&\left(\frac{2}{D}\right)^{\frac{D+2}{2D}}
(t-t_{coll})^{\frac{D+2}{2D}},\\ N_0(t)&=&\mu
\left(\frac{2}{D}\right)^{\frac{D}{2}}
(t-t_{coll})^{\frac{D}{2}}.\label{n0postt0}
\end{eqnarray}
We note the remarkable result that the central residual density
$\rho(0,t)=S_D^{-1}\rho_0(t)$ displays a universal behavior just
after $t_{coll}$, a result already obtained in \cite{cs1}.
Moreover, we find that $N_0(t)$ has the same form as the mass
found within a sphere of radius $r_0(t)$ below $t_{coll}$, given
in Eq.~(\ref{m0t0}).

Moreover, the scaling function $S$ satisfies
\begin{equation}
\frac{D+2}{D^2}\left(\frac{2D}{D+2}S+x S'\right)+ S(DS+xS')+\mu
x^{-D}(DS+xS'-1)=0. \label{scas2T0}
\end{equation}
The constant $\mu$ is determined by imposing that the large $r$ behavior of
$s(r,t)$ (or $\rho(r,t)$) exactly matches (not simply
proportional) that obtained below $t_{coll}$, which depends on the
shape of the initial condition as shown in \cite{cs1}.
Equation (\ref{scas2T0}) can be solved implicitly by looking for
solutions of the form $x^D=z[S(x)]$. After cumbersome but
straightforward calculations, we obtain the implicit form
\begin{equation}
1+\frac{x^D}{\mu}S(x)=\left[1 +\frac{x^D}{\mu}\left(
S(x)+\frac{2}{D^2}\right)\right]^{\frac{D}{D+2}},
\end{equation}
which coincides with the implicit solution given in \cite{cs1}.  Note
that $S(x)$ is a function of $x^D$. We check that the above result
indeed leads to $S(0)=D^{-1}$, and to the large $x$ asymptotics
\begin{equation}
S(x)\sim
\mu^{\frac{2}{D+2}}\,\left(\frac{2}{D^2}\right)^{\frac{D}{D+2}}
x^{-\frac{2D}{D+2}}.
\end{equation}

Note finally that for $T=0$, $N_0$ saturates to 1 in a finite
time, corresponding to the deterministic collapse of the outer
mass shell initially at $r=1$. Indeed, using Gauss' theorem, the
position of a particle initially at $r(t=0)=1$ satisfies
\begin{equation}
\frac{dr}{dt}=-r^{-(D-1)}.
\end{equation}
The position of the outer shell is then
\begin{equation}
r(t)=(1-Dt)^{1/D},
\end{equation}
which vanishes for $t_{end}=D^{-1}$.

\section{Post-collapse dynamics at $T>0$}
\label{sec_postT}

\subsection{Scaling regime}
\label{sec_sr}

In the more general case $T\ne 0$, we will proceed in a very
similar way as in the previous section. We define again,
\begin{equation}
s(r,t)=\frac{M(r,t)-N_0(t)}{r^D},
\end{equation}
where $N_0$ still satisfies
\begin{equation}
\frac{d N_0}{d t}=\rho_0 N_0. \label{N0Tn0}
\end{equation}
We now obtain
\begin{equation}
{\partial s\over\partial t}=T\left(\frac{\partial^2 s}{\partial
r^2}+\frac{D+1}{r}\frac{\partial s}{\partial r}\right) +\biggl
(r{\partial s\over\partial r}+Ds\biggr )s+\frac{N_0}{r^D}\biggl
(r{\partial s\over\partial r}+Ds-\rho_0\biggr ). \label{sTn0}
\end{equation}
By definition, we have again $s(0,t)=\rho_0(t)/D$.

We look for a scaling solution of the form
\begin{equation}
s(r,t)=\rho_{0}(t)S\biggl ({r\over r_{0}(t)}\biggr ),
\label{st0post}
\end{equation}
with $S(0)=D^{-1}$. As before,  we
define the King's radius by
\begin{equation}
r_0=\left(\frac{T}{\rho_0}\right)^{1/2}.
\end{equation}
For $t<t_{coll}$, we had $s(r,t)\sim 4Tr^{-2}$ (or $S(x)\sim
4x^{-2}$). In a very short time after $t_{coll}$, this property
should be preserved, which implies that the post-collapse scaling
function should also behave as
\begin{equation}
S(x)\sim 4x^{-2},\label{condas}
\end{equation}
for large $x$. Inserting the scaling $ansatz$ in Eq. (\ref{sTn0}), we obtain
\begin{equation}
{1\over 2\rho_0^{2}}\frac{d\rho_0}{dt}\left(2 S+x S'\right)= S''
+\frac{D+1}{x}S'+S(DS+xS')+
\frac{N_0}{\rho_0r_0^D}{1\over x^{D}}(DS+xS'-1). \label{scasTn0}
\end{equation}
Again, this equation should be time independent for scaling to
hold, which implies that there exist two constants $\mu$ and
$\kappa$ such that
\begin{equation}
N_0=\mu \rho_0r_0^D,
\end{equation}
and
\begin{equation}
\frac{1}{2\rho_0^{2}}\frac{d\rho_0}{dt}=-\kappa, \label{kappab}
\end{equation}
with $\kappa>0$, as the central residual density is again expected
to decrease.  Equation (\ref{kappab}) implies that $\rho_0\sim
(t-t_{coll})^{-1}$, and then that $N_0\sim (t-t_{coll})^{D/2-1}$.
We thus find a power law behavior for $N_0$, which in order to be
compatible with Eq.~(\ref{N0Tn0}), leads to the universal behavior
\begin{equation}
\rho_0(t)=\left(\frac{D}{2}-1\right)(t-t_{coll})^{-1},
\end{equation}
and then to
\begin{equation}
\kappa=\frac{1}{D-2}.
\end{equation}
We end up with the scaling equation
\begin{equation}
\frac{1}{D-2}\left(2 S+x S'\right)+S'' +\frac{D+1}{x}S'+S(DS+xS')+
\mu x^{-D}(DS+xS'-1)=0,\label{scasTn01}
\end{equation}
where $\mu$ has to be chosen so that $S(x)$ satisfies the condition of
Eq.~(\ref{condas}). Its value will be determined numerically in
Sec. \ref{sec_ns} for $D=3$.  Note that for small $x$, the
pre-collapse scaling function satisfies $S(x)-S(0)\sim x^2$, whereas
the post-collapse scaling function behaves as
\begin{equation}
S(x)-S(0)\sim x^D.\label{smallx}
\end{equation}
However, contrary to the $T=0$ case, $S(x)$ is not purely a
function of $x^D$.

Finally, we find that the weight of the central peak has a
universal behavior for short time after $t_{coll}$
\begin{equation}
N_0(t)=\mu\left(\frac{2}{D-2}\right)^{D/2-1}T^{D/2}\,(t-t_{coll})^{D/2-1}.
\label{n0post}
\end{equation}
Note that $N_0(t)$ behaves in a very similar manner to the mass
within a sphere of radius $r_0$ below $t_{coll}$, shown in
Eq.~(\ref{m0}). In addition, comparing Eq.~(\ref{n0post}) and
Eq.~(\ref{n0postt0}), we can define again a post-collapse
cross-over time between the $T\ne 0$ and $T=0$ regimes
\begin{equation}
t_*-t_{coll}\sim T^{D/2},
\end{equation}
which is similar to the definition of Eq.~(\ref{tst0}).

\subsection{Large time limit}
\label{sec_largetime}

Contrary to the $T=0$ case, the complete collapse does not take
place in a finite time as thermal fluctuations always allow for
some particle to escape the  central strongly attractive
potential. In order to illustrate this point, and obtain more
analytic insight on this matter, we will place ourselves in the
extreme situation where almost all the mass has collapsed
($N_0\approx 1$), and only an infinitesimal amount remains in
the residual profile.

In this limit, the residual density $\rho(r,t)$ satisfies the Fokker-Planck equation
\begin{equation}
\frac{\partial \rho}{\partial t}=T \left(\frac{\partial^2
\rho}{\partial r^2} +\frac{D-1}{r}\frac{\partial \rho}{\partial
r}\right) +{1\over r^{D-1}}\frac{\partial \rho}{\partial r},
\label{scalt}
\end{equation}
with boundary condition
\begin{equation}
T\frac{\partial \rho}{\partial r}(1,t)+\rho(1,t)=0.
\end{equation}
The problem indeed reduces to the study of a very light gas ({\it
i.e.}  with negligible self-interaction) of Brownian particles
submitted to the gravitational force ${\bf F}=-(GM/r^{D-1}){\bf
e}_{r}$ of a central unit mass. Alternatively, this can also be seen
as the probability distribution evolution equation of a system of two
Brownian particles moving in their mutual gravitation field.

Equation (\ref{scalt}) can be re-expressed as a Schr\"odinger equation (in
imaginary time), thus involving a self-adjoint operator (see Appendix
\ref{sec_schro}). The large time behavior is dominated by the first
eigenstate. Coming back to the notation of Eq.~(\ref{scalt}), we find
that
\begin{equation}
\rho(r,t)\sim {\rm e}^{-\lambda t}\psi(r),\label{decro}
\end{equation}
where $\psi$ satisfies the eigenequation
\begin{equation}
-\lambda \psi(r)=T \left(\psi'' +\frac{D-1}{r}\psi'\right)
+{1\over r^{D-1}}\psi',\label{eigen1}
\end{equation}
and the same boundary condition as $\rho$, i.e.
\begin{equation}
T\psi'(1)+\psi(1)=0.
\label{eigenb}
\end{equation}
The eigenvalue $\lambda$ will also control the large time behavior of $\rho_0$ and $N_0$
as Eq.~(\ref{N0Tn0}) and Eq.~(\ref{decro}) both imply that
\begin{equation}
1-N_0(t)\approx \frac{\rho_0(t)}{\lambda}\sim {\rm e}^{-\lambda t}.
\label{decexp}
\end{equation}

We did not succeed in  solving analytically the above eigenequation,
and for a given temperature, this has to be solved numerically.
However, in the limit of very small temperature, we can apply
techniques reminiscent from semiclassical analysis in quantum
mechanics ($T\leftrightarrow\hbar$). We now assume $T$ very small
and define $\phi$ such that
\begin{equation}
\psi(r)={\rm e}^{-\frac{\phi(r)}{T}}.
\label{change}
\end{equation}
The function $h=\phi'$ satisfies the following non-linear first order
differential equation
\begin{equation}
T\left(h'+\frac{D-1}{r}h\right)+\frac{h}{r^{D-1}}-h^2=\lambda
T,
\label{heq}
\end{equation}
with the simple boundary condition
\begin{equation}
h(1)=1.\label{cond}
\end{equation}

In the limit $T\to 0$, the term proportional to $T$ in the
left-hand side of Eq.~(\ref{heq}) can {\it a priori} be discarded
leading to
\begin{equation}
h(r)=\frac{2\lambda Tr^{D-1}}{1+\sqrt{1-4\lambda
Tr^{2(D-1)}}}.\label{h0}
\end{equation}
If $4\lambda T<1$, the above expression is a valid perturbative
solution also at $r=1$, but cannot satisfies the constraint
$h(1)=1$. Hence, we conclude that in the limit of small
temperature $4\lambda T\geq 1$, so that the above expression is
only valid for $r$ not to close to $r=1$. The above argument also
suggests that $\lambda T$ is of order unity and we write
\begin{equation}
\lambda T=\frac{1}{4}+\mu^2.
\end{equation}
To understand how the boundary condition Eq.~(\ref{cond}) can be
in fact satisfied, one has to come back to Eq.~(\ref{heq}), which
for $r=1$, shows that $h'(1)\sim\lambda\sim T^{-1}\gg 1$. This
implies that the term $Th'$ cannot be neglected near $r=1$, and
that $h$ varies in a noticeable way on a length scale from 1 of
order $T$.

This suggests to define
\begin{equation}
z(x)=h(1-Tx)
\end{equation}
which satisfies (at order 0 in $T$)
\begin{equation}
-z'+z-z^2=\frac{1}{4}+\mu^2,
\end{equation}
and $z(0)=1$. This equation has the unique solution
\begin{equation}
z(x)=\frac{1}{2}+\mu\frac{1-2\mu\tan(\mu x)}{2\mu+\tan(\mu x)}.
\end{equation}
For large $x$, this function only has a sensible behavior for
$\mu=0$, which shows that
\begin{equation}
\lim_{T\to 0}\lambda T=\frac{1}{4},
\end{equation}
and that Eq.~(\ref{h0}) is in fact valid for $1-r\gg (4\lambda
T-1)\to 0$. To leading order, we find
\begin{equation}
h(1-Tx)\approx z_0(x)=\frac{1}{2}+\frac{1}{2+x}.\label{z0}
\end{equation}

Equations (\ref{h0}) and (\ref{z0}), show that $h(x)$ goes rapidly
from 1 to $1/2$ in a small region close to $r=1$ where $h$ varies
on the scale $T$. One can even compute the next correction to
$\lambda T$ by including the next term of order $T$ in the
equation for $z$. Writing
\begin{equation}
z(x)=z_0(x)+T^{1/3}z_1(xT^{1/3}), \label{z1}
\end{equation}
we find
\begin{equation}
z_1'+\frac{2}{u}z_1+z_1^2-\frac{D-1}{2}u=-\frac{\mu^2}{T^{2/3}}=-c_D,\qquad
u=T^{1/3}x.\label{eigenz1}
\end{equation}
This is again an eigenvalue problem which selects a unique
constant $c_D$, that we could only solve numerically. Still, this
leads to the non trivial result
\begin{equation}
\lambda=\frac{1}{4T}+\frac{c_D}{T^{1/3}}+...\qquad (T\rightarrow 0).
\label{devl}
\end{equation}

We now solve the eigenvalue problem (\ref{eigen1}) (\ref{eigenb}) in
the limit of large temperatures $T\rightarrow +\infty$ (see also Appendix \ref{sec_schro}).  We again perform the change of variables (\ref{change}) and
rewrite Eq. (\ref{heq}) in the form
\begin{equation}
h'+\frac{D-1}{r}h=\lambda-{1\over T}\biggl ({h\over r^{D-1}}-h^{2}\biggr ).
\label{add1}
\end{equation}
Then, we expand the solutions of this equation in terms of the small parameter $1/T\ll 1$. We write $h=h_{0}+{1\over T}h_{1}+{1\over T^{2}}h_{2}+...$ and $\lambda=\lambda_{0}+{1\over T}\lambda_{1}+{1\over T^{2}}\lambda_{2}+...$. To zeroth order, we have
\begin{equation}
h_{0}'+\frac{D-1}{r}h_{0}=\lambda_{0}.
\label{add2}
\end{equation}
The solution of this equation is $h_{0}=\lambda_{0}r/D$. Using the boundary condition $h_{0}(1)=1$, we obtain
\begin{equation}
\lambda_{0}=D, \qquad h_{0}=r,
\label{add3}
\end{equation}
and $h_{n}(1)=0$ for $n>0$. To first order, we get
\begin{equation}
h_{1}'+\frac{D-1}{r}h_{1}=\lambda_{1}-{h_{0}\over r^{D-1}}+h_{0}^{2}.
\label{add4}
\end{equation}Integrating this first order differential equation and using the boundary condition $h_{1}(1)=0$, we obtain
\begin{equation}
h_{1}={D\over 2(D+2)}r-{1\over 2}r^{3-D}+{r^{3}\over D+2},
\label{add5}
\end{equation}
with
\begin{equation}
\lambda_{1}={D^{2}\over 2(D+2)}.
\label{add5b}
\end{equation}
Hence, the large temperature behavior of the eigenvalue is
\begin{equation}
\lambda=D+{D^{2}\over 2(D+2)}{1\over T}+...\qquad (T\rightarrow +\infty).
\label{add5c}
\end{equation}
This expansion can be easily carried out to higher orders but the coefficients are more and more complicated. Restricting ourselves to $D=3$, we get $\lambda=3+{9\over 10}{1\over T}-{477\over 700}{1\over T^{2}}+O(T^{-3})$.

\section{Numerical simulations in the canonical ensemble}
\label{sec_ns}

In this section, we illustrate the analytical results obtained in the
previous section in the case of $D=3$. Except when specified
otherwise, our simulations have been performed at $T=1/5<T_c\approx
0.397...$, for which we have obtained $t_{coll}\approx
0.44408...$.

\begin{figure}
\centerline{
\psfig{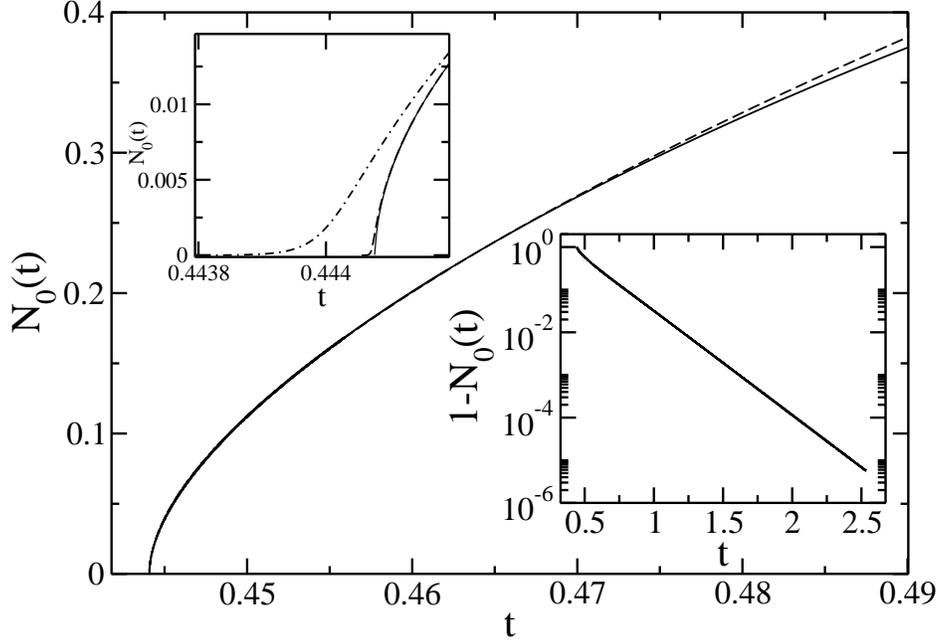}} \vskip
0.2cm \caption{We plot $N_0(t)$ for small times (full line). This
is compared to $N_0(t)^{\rm
Theory}{\times}\left[1+a(t-t_{coll})^b\right]$ (dashed line), where
$N_0(t)^{\rm Theory}$ is given by Eq.~(\ref{n0post}) with
$\mu=8.38917147...$, and $a\approx 1.7.$ and $b\approx 0.33$
are fitting parameters. Note that the validity range of this fit
goes well beyond the estimated $t_*$ with $t_*-t_{coll}\sim
T^{D/2}\sim 0.09$. The bottom insert illustrates the exponential
decay of $1-N_0(t)\sim{\rm e}^{-\lambda t}$. The best fit for
$\lambda$ leads to $\lambda\approx 5.6362$ to be compared to the
eigenvalue computed by means of Eq.~(\ref{eigen1}), $\lambda=
5.6361253...$. Finally, the top inset illustrates the
sensitivity of $N_0(t)$ to the space discretization, which
introduces an effective cut-off (a factor 4 between each of the 3
curves). Note the small time scale. Even the curve corresponding
to the coarsest discretization becomes indistinguishable from the
others for $t>0.448$.} \label{fig1}
\end{figure}

\begin{figure}
\centerline{
\psfig{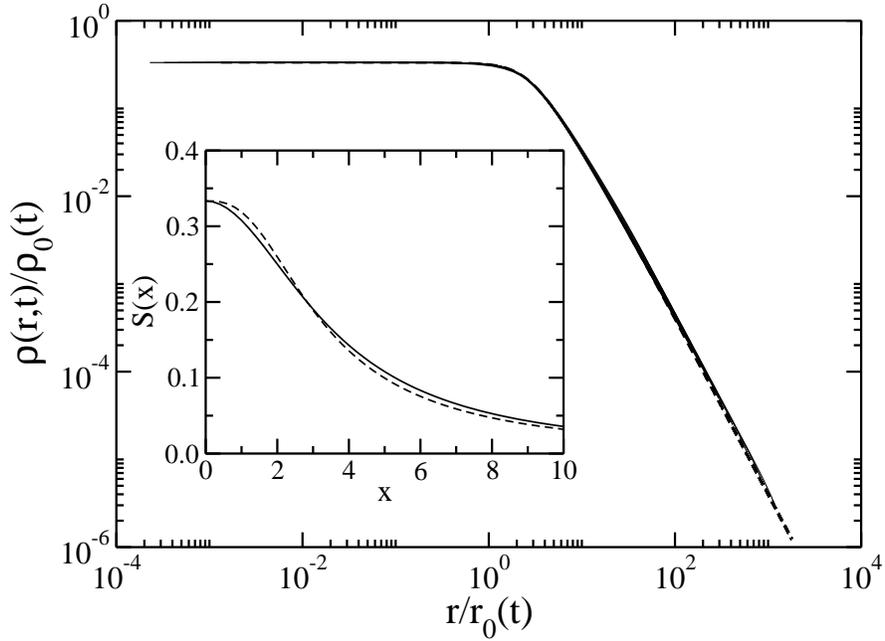}}
\vskip 0.2cm
\caption{In the post-collapse regime, we plot $\rho(r,t)/\rho_0(t)$ as
a function of the scaling variable $x=r/r_0(t)$. A good data collapse
is obtained for central residual densities in the range $10^3\sim
10^6$. This is compared to the numerical scaling function computed
from Eq.~(\ref{scasTn01}) (dashed line). The insert shows the
comparison between this post-collapse scaling function (dashed line)
and the scaling function below $t_{coll}$ which has been rescaled to
have the same value at $x=0$, preserving the asymptotics:
$S(x)=(3+x^2/4)^{-1}$ (see Eq.~(\ref{solscad}); full line). Note that
the post-collapse scaling function is flatter near $x=0$, as
$S(x)-1/3\sim x^3$ (in $D=3$) above $t_{coll}$ instead of
$S(x)-1/3\sim x^2$, below $t_{coll}$. }
\label{fig2}
\end{figure}

\begin{figure}
\centerline{
\psfig{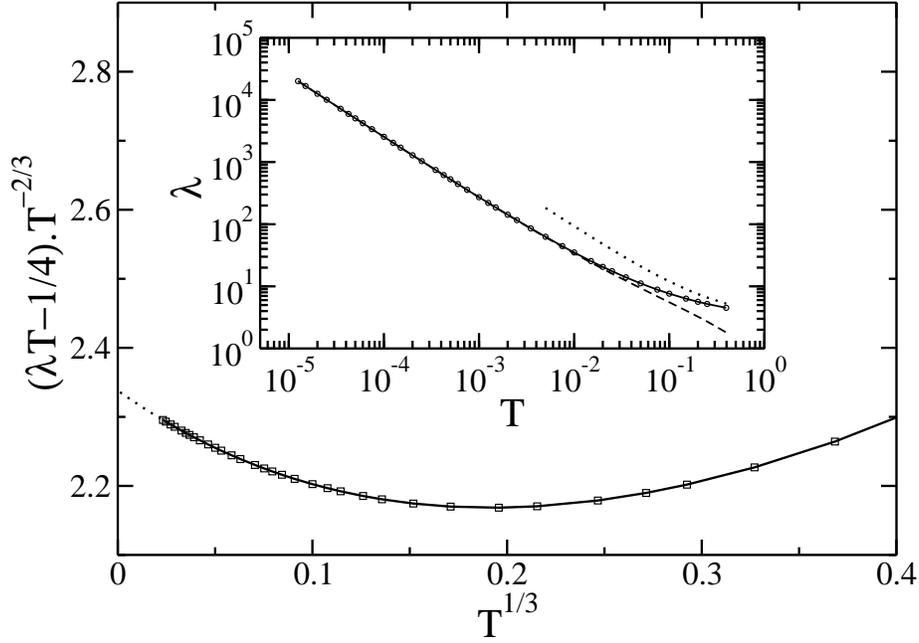}}
\vskip 0.2cm
\caption{The main plot represents 
$\left(\lambda T-\frac{1}{4}\right) T^{-2/3}$ as a function of
$T^{1/3}$ (line and squares), which should converge to $c_{D=3}=
2.33810741...$ for $T\rightarrow 0$ according to
Eq.~(\ref{devl}). We find a perfect agreement with this value using a
quadratic fit (dotted line).  Furthermore, this fit shows that the
slope at $T=0$ is in fact equal to $-2\pm 2.10^{-4}$, suggesting that
the next term to the expansion of Eq.~(\ref{devl}) is
$\lambda=\frac{1}{4T}+\frac{c_3}{T^{1/3}}-2+...$, in $D=3$. In the
insert, we plot $\lambda$ as a function of $T$ up to $T\approx
T_{c}\approx 0.4$. The small temperature analytical result of
Eq.~(\ref{devl}) is in very good agreement with the numerical data up
to $T\sim 0.03$, whereas the large $T$ estimate
$\lambda(T)=3+\frac{9}{10}T^{-1}+...$ is only qualitatively correct
in this range of physical temperatures $T\leq T_{c}$.  }
\label{fig3}
\end{figure}

\begin{figure}
\centerline{
\psfig{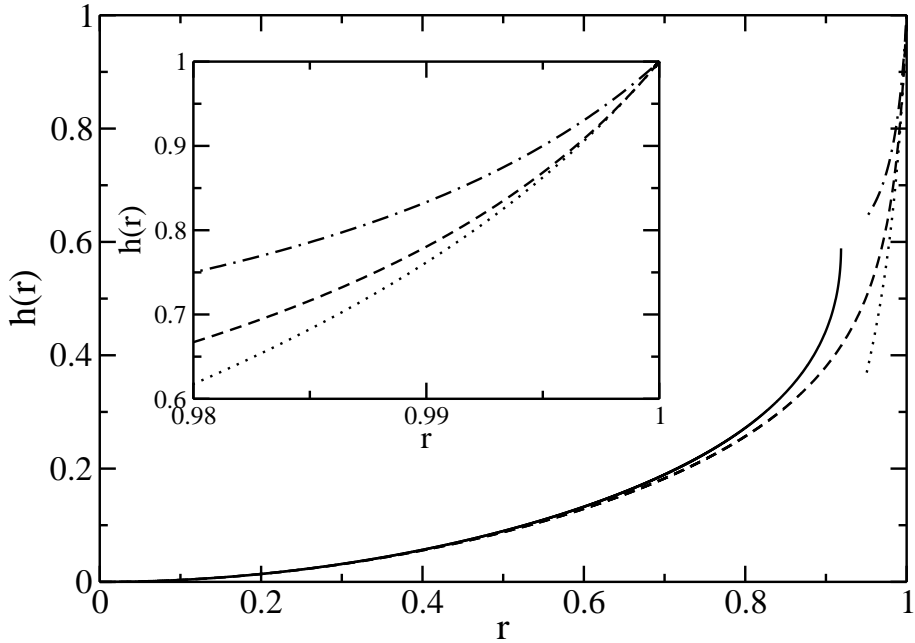}}
\vskip 0.2cm
\caption{For $T=0.01$ ($\lambda=35.074198...$, $\mu=
0.31739877...$), we plot $h(r)$ computed numerically from
Eq.~(\ref{heq}) (dashed line), and the theoretical expression of
Eq.~(\ref{h0}), which is valid for $1-r\gg \mu^2\sim T^{2/3}$ (full
line).  We also plot the theoretical expression of $z_0$ (dashed-dot
line; see Eq.~(\ref{z0})) and the next order perturbation result (dot line; 
see Eq.~(\ref{z1})), which are valid in the region $1-r\ll \mu^2\sim
T^{2/3}$. The insert is a blow up of the region close to 1. Note how
$h(r)$ varies by a quantity of order unity as $r$ varies by a quantity
of order $T=0.01$. We have chosen a not too small value for $T$ in
order to be able to visualize the two scale regimes on a single
figure. Both approximations shown in the insert are getting better as
$T$ decreases.}
\label{fig4}
\end{figure}

In order to perform our simulations, we have used a Runge-Kutta
algorithm with adaptive step in space and time. We call $dr$ the
spatial discretization near $r=0$ (which we need to take very small
as the density profile becomes singular at $r=0$). An important
numerical problem arises in the numerical integration of
Eq.~(\ref{N0Tn0}), which is crucial in obtaining non zero values for
$N_0(t)$. As this equation is a first order differential equation with
initial condition $N_0(0)=0$, any naive integration scheme should lead
to a strictly vanishing value for $N_0(t)$ for all time, and any $dr$. Still,
when performing this naive numerical integration, we see that crossing
$t_{coll}$ generates increasing values for $M(dr,t)$, although
keeping $M(0,t)=N_0(t)=0$ ultimately makes the numerical integration
unstable. In order to bypass this problem, we have decided to introduce
a numerical scheme where Eq.~(\ref{N0Tn0}) is replaced by
\begin{equation}
\frac{d N_0}{d t}=\rho_0^{fit} N_0^{fit}. \label{N0Tn0new}
\end{equation}
$N_0^{fit}$ and $\rho_0^{fit}$ are extracted from a fit of $M(r,t)$ to
the functional form (we are in $D=3$)
\begin{equation}
M(r,t)\approx
N_0^{fit}(t)+\frac{\rho_0^{fit}(t)}{3}r^3+a_5(t)r^5+a_6(t)r^6,
\end{equation}
in a region of a few $dr$, excluding of course $r=0$. This functional
form is fully compatible with the expected expansion for $M(r,t)$,
both below ($a_6=0$) and above ($a_5=0$) $t_{coll}$. We find that this
numerical scheme allows us to cross smoothly the singularity at
$t_{coll}$. An effective cut-off is introduced which effectively
depends on $dr$, and we have checked that the result presented in this
section are extremely close to the ones that would be obtained in the
ideal limit $dr\to 0$. This is  illustrated in Fig.~1, where the smoothing
effect of our algorithm is shown to act on a very small time region
after $t_{coll}$. Even more surprisingly, we find that for
sufficiently large times (actually very small compared to any physical
time scales), our results are essentially independent of $dr$, even
for unreasonably large values of $dr$. We are thus confident that we
have successfully crossed the collapse singularity.

In Fig.~1, we plot $N_0(t)$ for small time which compares well with
the universal form of Eq.~(\ref{n0post}), where $\mu=8.38917147...$
has been determined so as to ensure the proper behavior of $S(x)$ for
large $x$ (see Sec. \ref{sec_sr}).  We also illustrate the exponential
decay of $1-N_0(t)$, with a rate in perfect agreement with the value
of $\lambda$ extracted from solving numerically the eigenvalue problem
of Eq.~(\ref{eigen1}). Finally, we show the effect of the numerical
spatial discretization $dr$ near $r=0$. Satisfactorily enough, the
value of $N_0(t)$ is sensitive to the choice of $dr$ only for very
small times after the collapse, and we were able to easily reach small
enough $dr$, in order to faithfully reproduce the post-collapse
singularity. In Fig.~2, we convincingly illustrate the post-collapse
scaling, and compare the post-collapse scaling function to that
obtained analytically below $t_{coll}$ (pre-collapse). In Fig.~3, we
confirm the validity of our perturbative expansion for $\lambda$, in
the limit of small temperature. We compare the value of $c_D$
extracted from directly solving the full eigenvalue problem to that
obtained from Eq.~(\ref{eigenz1}), finding a perfect
agreement. Finally, in Fig.~4, we compare the numerical value obtained
for $h(r)$ to the different analytical estimates given in the
preceding section, for $T=0.01$. The two important regions $1-r\ll
\mu^2\sim T^{2/3}$ and $1-r\gg \mu^2\sim T^{2/3}$ can be clearly
identified.

\section{Post-collapse in the microcanonical ensemble}
\label{sec_postmicro}

So far, we have only addressed the post-collapse dynamics in the
canonical ensemble. In the microcanonical ensemble, the dynamical
equation have to be supplemented with the strict energy conservation
condition (see Eq.~(\ref{dim3})), which fixes the global temperature
$T(t)$. For this model, it was shown in \cite{charosi,cs1,cs2} that
below a certain energy (Antonov energy), the system collapses with an
apparent scaling associated with $\alpha_{max}\approx 2.2$ for
intermediate times (when the temperature still increases in a
noticeable way) before entering a scaling regime with $\alpha=2$,
identical to that obtained in the canonical ensemble. In the limit
$t\to t_{coll}$, the temperature and the potential energy both seem to
converge to a finite value preserving a constant energy. Closely
before the collapse time, the temperature behaves as
$T(t_{coll})-T(t)\sim (t_{coll}-t)^{\gamma}$ with $\gamma\simeq
1/2$. This section addresses the $t>t_{coll}$ time period.

Assuming a spherical mass density, and after integration by parts,
the potential energy $W$ can be rewritten in the form ($D>2$)
\begin{equation}
W(t)=-\frac{1}{2}\int_0^{1}\frac{M^2(r,t)}{r^{D-1}}\,dr
-\frac{1}{2(D-2)}.
\end{equation}
We see immediately that as $D-1>1$, the occurrence of a finite mass
$N_0(t)\ne 0$ concentrated at $r=0$ implies an infinite potential
energy, hence an infinite temperature. We thus anticipate that the
post-collapse dynamics in the microcanonical ensemble is probably an
ill-defined problem. In this extreme regime, let us try to consider
the possible flaws of this model in order to describe a consistent
dynamics of a reasonable physical self-gravitating system. First, our
assumption of uniform temperature is certainly not realistic in a
system displaying huge density contrast, and some alternative
approaches are needed to incorporate a spatially dependent temperature. This
point is certainly crucial and will be addressed in a future work
\cite{cs3}. Furthermore, in this regime, a careful physical analysis
predicts that this system of self-gravitating individual particles
should lead to the formation of binaries which is probably beyond the
description ability of our essentially mean-field approach. In other
words, the system may become intrinsically heterogeneous, which
probably cannot be captured by our continuous model. Finally, we can
think of other physical effects (degeneracy effects of quantum or
dynamical origin, finite particle size effects,...) preventing the
system from reaching arbitrarily large densities. One way to describe
such effects consists in introducing a spatial cut-off $h$ or a
density cut-off of order $h^{-D}$. In such a system, the dynamics
first follows the pre-collapse dynamics until the maximum density is
approached.  Then, the system will ultimately reach a maximum entropy
state that we propose to characterize in a simple manner, as in
\cite{cs,fermions}.

We propose to describe the final state as a ``core-halo'' structure,  
which for simplicity we modelize as a core of radius $h\ll 1$ and
constant density
\begin{equation}
\rho_{core}=\frac{DN_0}{S_Dh^D},
\end{equation}
which mimics a regularized central Dirac peak containing a
mass $N_0$. In the region $h<r\leq 1$ stands the halo of constant
density
\begin{equation}
\rho_{halo}=\frac{D(1-N_0)}{S_D(1-h^D)},
\end{equation}
containing the rest of the mass. As $h$ is small, we can compute
the potential (or total) energy and the entropy only including the
relevant leading terms. We find
\begin{equation}
E=\frac{D}{2}T-\frac{D}{D^2-4}\left[\frac{N_0^2}{h^{D-2}}+
1+\frac{D-2}{2}N_0-\frac{D}{2}N_0^2\right]+O(h^2),
\end{equation}
where the first term is the kinetic energy, whereas the entropy
(up to irrelevant constants) reads
\begin{equation}
S=\frac{D}{2}\ln T-N_0\ln\left(\frac{N_0}{h^D}\right)
-(1-N_0)\ln(1-N_0)+O(h^D).
\end{equation}
For a given small value of $h$, $S$ has a local maximum at $N_0=0$
provided that $E>E_c(h)$, with $\lim_{h\to
0}E_c(h)=-\frac{D}{D^2-4}$. Below $E_c$, the sole entropy maximum
resides at $N_0$ satisfying the following implicit equation (again
in the limit of small $h$)
\begin{equation}
N_0=\frac{D}{\ln\left(\frac{N_0}{h^D}\right)}\sim -\frac{1}{\ln
h},\label{N0h}
\end{equation}
where the given asymptotics is quantitatively correct only for
extremely small values of $h$. Hence, we find that the mass included
in the core slowly {\it decreases} with the core size $h$
\cite{fermions}, resulting in an effective singularity $\rho({\bf
r})=-{1\over\ln r}\delta^{D}({\bf r})$. Meanwhile, the temperature
diverges as
\begin{equation}
T\sim-\frac{2}{D}W\sim\frac{2}{D^2-4} \frac{1}{h^{D-2}\ln^2
h}.
\end{equation}
and leads rapidly and efficiently to a uniform halo.

\begin{figure}
\centerline{
\psfig{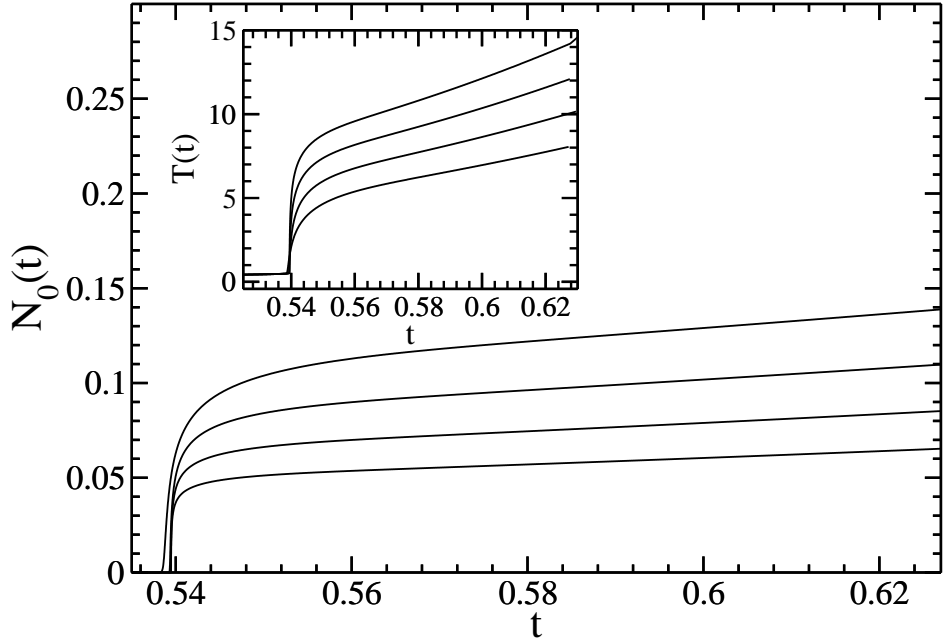}
}
\vskip 0.2cm \caption{For $E=-0.45<E_c\approx-0.335$, we plot
$N_0(t)$, for different values of $h$ decreasing by a factor 2 for
each curve from the top one to the bottom one. It is clear that
$N_0(t)$ decreases as $h$ decreases. The insert shows the
corresponding temperature plots. The temperature $T(t)$ increases as $h$
decreases. Note that in the pre-collapse regime, the temperature
essentially does not depend on $h$. For the situation considered, it
starts from $T(0)=0.1$ and culminates at $T(t_{coll})\approx 0.5$. }
\label{fig5}
\end{figure}

In order to relate this result to our actual system, we perform
microcanonical post-collapse simulations using the same
regularization scheme as in the canonical case in order to
describe the evolution of $N_0(t)$. In addition to this, we also
need to regularize the potential energy which is strictly infinite
(as well as $T$) when $N_0\ne 0$. Consistently with the previous
discussion, we introduce a numerical cut-off $h$ by defining
\begin{equation}
W(t)=-\frac{1}{2}\int_h^{1}\frac{M^2(r,t)}{r^{D-1}}\,dr
-\frac{1}{2(D-2)}.
\end{equation}
In Fig.~5, we plot $N_0(t)$ and $T(t)$ for different values of the
cut-off $h$. Contrary to the canonical case, the post-collapse
dynamics is strongly $h$ dependent. We see that in conformity with our
result of Eq.~(\ref{N0h}), the central mass $N_0$ clearly decreases as
$h\to 0$. Therefore, in the microcanonical case, the physical picture
is that when the collapse time $t_{coll}$ is reached, the temperature
increases rapidly, which leads to the rapid homogenization of the
system except for a dense and small core, whose mass  $N_0\sim
\ln^{-1}T$ is a decreasing function of the maximum temperature reached.
This central structure with weak mass and huge binding energy is
similar to a ``binary star'' structure in stellar dynamics. Binary
formation is the physical process that arrests core collapse in
globular clusters \cite{henon}. This is also the end point of our
simple microcanonical Brownian model.

\section{Conclusion}
\label{sec_conclusion}

In this paper, we have investigated the post-collapse dynamics of a
gas of self-gravitating Brownian particles in canonical and
microcanonical ensembles. Our results also apply to the chemotactic
aggregation of bacterial populations in biology. At the collapse time
$t_{coll}$, the system develops a singular density profile scaling as
$\rho\sim r^{-2}$. However, the ``central singularity contains no
mass'', the temperature does not diverge, and the entropy and free
energy are finite \cite{charosi,cs1,cs2}. Since this profile is not a
maximum entropy (resp. minimum free energy) state nor a stationary
solution of the Smoluchowski-Poisson system, the collapse continues
after $t_{coll}$.  This solves the apparent paradox reported in
\cite{fermions}.

In the canonical ensemble, mass accretes progressively at the center
of the system and a Dirac peak forms by swallowing the surrounding
particles. Eventually, the Dirac peak contains all the mass. This
structure has an infinite free energy $F=E-TS\rightarrow -\infty$
simply because its binding energy is infinite. This is therefore the
most probable structure in canonical ensemble
\cite{kiessling,chavcano,fermions}. In the microcanonical ensemble,
the maximum entropy state (at fixed mass and energy) consists of a
single binary embedded in a hot halo
\cite{antonov,paddy,cs1}. This is precisely what we see in our
numerical simulations. The temperature increases dramatically above
$t_{coll}$ (resulting in an almost uniform halo) although the mass
contained in the core is weak (but finite). We note the
``spectacular'' fact that almost all the gravitational energy resides
in a binary-like core with negligible mass. A similar phenomenon is
observed in stellar dynamics for globular clusters having experienced
core collapse \cite{bt}. This shows that the microcanonical
Smoluchowski-Poisson system shares some common properties with kinetic
equations usually considered in stellar dynamics (Landau-Fokker-Planck
equations), despite its greater simplicity. Clearly, a major drawback
of our microcanonical model is to assume that the temperature
uniformizes instantaneously, implying an infinite thermal
conductivity. We shall relax this simplification in a future work
\cite{cs3}. However, the present study is one of the first {dynamical}
study showing the formation of Dirac peaks and binary-like structures
in systems with gravitational interaction.

Our Brownian model is based on the existence of a gaseous medium that
generates a friction force. This situation exists in certain
astrophysical models such as the transport of dust particles in the
solar nebula \cite{planete}. Dust particles are submitted to Stokes or
Epstein drag. It is clear that when the concentration of particles is
important (prior to planetesimal formation), self-gravity has to be taken
into account. Thus, our system of self-gravitating Brownian particles
could be connected to this astrophysical situation.  We just mention
this as a possible astrophysical application because it is not our
present main motivation to make a precise model of dust-gas-gravity
coupling in protoplanetary disks. However, this problem could be
considered in future works.

\newpage
\appendix

\section{Brownian particle around a ``black hole''}
\label{sec_schro}

We consider the Brownian motion of a particle subject to the
gravitational force $-{GM\bf r}/r^{3}$ created by a central mass $M$
(``black hole''). We assume that when the particle comes at $r=0$, it
is captured by the central mass. We denote by $W({\bf r},t)$ the
density probability of finding the particle in ${\bf r}$ at time
$t$. It is solution of the Fokker-Planck equation
\begin{equation}
{\partial W\over\partial t}=\nabla\biggl (T\nabla W+W{{\bf r}\over r^{3}}\biggr ),
\label{devl1}
\end{equation}
where we have set $G=M=R=\xi=1$. Let $W(r,t)$ denote a spherically symmetric
solution of Eq. (\ref{devl1}) satisfying the boundary conditions
\begin{equation}
T{\partial W\over\partial r}(1,t)+W(1,t)=0,
\label{devl2}
\end{equation}
\begin{equation}
W(r,0)={\delta(r-r_{0})\over 4\pi r_{0}^{2}}.
\label{devl3}
\end{equation}
We call
\begin{equation}
{\bf J}=-\biggl (T\nabla W+W{{\bf r}\over r^{3}}\biggr ),
\label{devl4}
\end{equation} 
the current of probability, i.e. ${\bf J}dS {\bf n}$ gives the
probability that the particle crosses an element of surface $dS$ between
$t$ and $t+dt$ (${\bf n}$ is a unit vector normal to the element of
surface under consideration).

We introduce the probability $p(r_{0},t)dt$ that a particle
located initially between $r_{0}$ and $r_{0}+dr_{0}$ arrives for the first
time at $r=0$ between $t$ and $t+dt$. We have
\begin{equation}
p(r_{0},t)=-\int_{R_{\epsilon}}{\bf J}\cdot d{\bf S}= 4\pi\epsilon^{2}\biggl
(T{\partial W\over\partial r}+{W\over r^{2}}\biggr )_{\epsilon}=4\pi
W(0,t),
\label{devl5}
\end{equation} 
where $R_{\epsilon}$ is a ball of radius $\epsilon\rightarrow 0$. The
total probability that the particle initially between $r_{0}$ and
$r_{0}+dr_{0}$ has reached the center of the system between $0$ and $t$ is 
thus
\begin{equation}
Q(r_{0},t)=\int_{0}^{t}p(r_{0},t')dt'.
\label{devl6}
\end{equation} 
Finally, we average $Q(r_{0},t)$ over an appropriate range of initial positions in order to get the expectation $Q(t)$ that the particle has been captured at time $t$. 

With the change of variables 
\begin{equation}
W=\psi e^{1\over 2Tr},
\label{devl7}
\end{equation} 
we can transform the Fokker-Planck equation (\ref{devl1}) into a Schr\"odinger equation (in imaginary time) of the form
\begin{equation}
{\partial\psi\over\partial t}=T\Delta\psi-{1\over 4Tr^{4}}\psi.
\label{devl8}
\end{equation} 
A separation of the variables can be effected by the substitution
\begin{equation}
\psi=\phi(r)e^{-\lambda t}.
\label{devl9}
\end{equation} 
This transformation reduces the Schr\"odinger equation to a second order ordinary differential equation
\begin{equation}
\phi''+{2\over r}\phi'+\biggl ({\lambda\over T}-{1\over 4T^{2}r^{4}}\biggr )\phi=0,
\label{devl10}
\end{equation} 
with the boundary condition
\begin{equation}
\phi'(1)+{1\over 2T}\phi(1)=0.
\label{devl11}
\end{equation} 
We note $\lambda_{n}$ the eigenvalues and $\phi_{n}$ the corresponding
eigenfunctions. Since the Schr\"odinger operator
$H=\Delta-1/4T^{2}r^{4}$ is Hermitian, the eigenfunctions form a complete set of orthogonal functions for the scalar product
\begin{equation}
\langle f g \rangle=\int_{0}^{1}f(r)g(r)4\pi r^{2}dr.
\label{devl12}
\end{equation} 
The system can be furthermore  normalized,
i.e. $\langle\phi_{n}\phi_{m}\rangle=\delta_{nm}$. Any function $f(r)$ satisfying the boundary condition (\ref{devl11}) can be expanded on this basis, as
\begin{equation}
f(r)=\sum_{n}\langle f\phi_{n}\rangle \phi_{n}.
\label{devl13}
\end{equation} 
In particular,
\begin{equation}
{\delta (r-r_{0})\over 4\pi r_{0}^{2}}=\sum_{n}\phi_{n}(r_{0})\phi_{n}(r).
\label{devl14}
\end{equation} 

The general solution of the problem (\ref{devl1}) (\ref{devl2}) can be
expressed in the form
\begin{equation}
W(r,t)=\sum_{n}A_{n}e^{-\lambda_{n}t}e^{1\over 2Tr}\phi_{n}(r),
\label{devl15}
\end{equation} 
where the coefficients $A_{n}$ are determined by the initial conditions (\ref{devl3}), using the expansion (\ref{devl14}) for the $\delta$-function. We get
\begin{equation}
W(r,t)=e^{{1\over 2T} ({1\over r}-{1\over r_{0}} )}\sum_{n} e^{-\lambda_{n}t}\phi_{n}(r_{0})\phi_{n}(r).
\label{devl16}
\end{equation} 
From this expression, we obtain
\begin{equation}
p(r_{0},t)=4\pi e^{-{1\over 2T r_{0}} }\sum_{n} e^{-\lambda_{n}t}\phi_{n}(r_{0})\lim_{r\rightarrow 0}\biggl\lbrack \phi_{n}(r)e^{1\over 2Tr}\biggr\rbrack.
\label{devl17}
\end{equation} 
Then, according to Eq. (\ref{devl6}), we have
\begin{equation}
Q(r_{0},t)=4\pi e^{-{1\over 2T r_{0}} }\sum_{n} {1-e^{-\lambda_{n}t}\over \lambda_{n}}\phi_{n}(r_{0})\lim_{r\rightarrow 0}\biggl\lbrack \phi_{n}(r)e^{1\over 2Tr}\biggr\rbrack.
\label{devl18}
\end{equation} 
Finally, averaging over the initial conditions, the probability that the particle has been captured by the central mass at time $t$ can be expressed as
\begin{equation}
Q(t)=\sum_{n} Q_{n}(t),
\label{devl19}
\end{equation} 
where 
\begin{equation}
Q_{n}(t)=B_{n} (1-e^{-\lambda_{n}t}), 
\label{devl20}
\end{equation} 
and
\begin{equation}
B_{n}=4\pi {1\over \lambda_{n}} \overline{e^{-{1\over 2T r_{0}} } \phi_{n}(r_{0})}\lim_{r\rightarrow 0}\biggl\lbrack \phi_{n}(r)e^{1\over 2Tr}\biggr\rbrack.
\label{devl21}
\end{equation} 
This formally solves the problem. If we consider the large time limit, we just need to determine the first eigenvalue $\lambda_{0}(T)$ of the quantum problem. This has been done analytically in Sec. \ref{sec_largetime} in the limits $T\rightarrow 0$ and $T\rightarrow +\infty$. 

Below, we consider again the high temperature regime where thermal
fluctuations prevail over gravity but we do not restrict ourselves to
the first eigenvalue. To leading order in the limit $T\rightarrow
+\infty$, the Fokker-Planck equation (\ref{devl1}) reduces to the pure
diffusion equation
\begin{equation}
{\partial W\over\partial t}=T{1\over r^{2}}{\partial\over\partial r}\biggl (r^{2}{\partial W\over\partial r}\biggr ).
\label{devl22}
\end{equation}
However, for consistency (see Sec. \ref{sec_largetime}), it is necessary to keep the term of order $1/T$ (arising from the gravitational force) in the boundary condition. Hence, we take 
\begin{equation}
{\partial W\over\partial r}(1,t)+{1\over T}W(1,t)=0.
\label{devl22b}
\end{equation}
The general solution of the diffusion equation (\ref{devl22}) with the boundary conditions
(\ref{devl22b}) and (\ref{devl3}) can be expressed as
\begin{equation}
W(r,t)=\sum_{n}e^{-\lambda_{n}t}\phi_{n}(r_{0})\phi_{n}(r),
\label{devl23}
\end{equation}
where $\phi$ is solution of 
\begin{equation}
\phi''+{2\over r}\phi'+{\lambda\over T}\phi=0,
\label{devl24}
\end{equation}
\begin{equation}
\phi'(1)+{1\over T}\phi(1)=0.
\label{devl25}
\end{equation}
Setting $\phi=\chi/r$, Eqs. (\ref{devl24}) and (\ref{devl25}) become
\begin{equation}
\chi''+{\lambda\over T}\chi=0,
\label{devl26}
\end{equation}
\begin{equation}
\chi'(1)=\biggl (1-{1\over T}\biggr )\chi(1).
\label{devl27}
\end{equation}
Equation (\ref{devl26}) is readily solved. The eigenvalues can be written $\lambda_{n}=Tx_{n}^{2}(T)$, where $x_{n}(T)$ are the solutions of the implicit equation
\begin{equation}
\tan(x_{n})={x_{n}\over 1-{1\over 2T}}.
\label{devl28}
\end{equation}
The eigenfunctions are
\begin{equation}
\phi_{n}(r)=A_{n}{\sin(x_{n}r)\over r}.
\label{devl29}
\end{equation}
The general solution of the diffusion equation can thus be written
\begin{equation}
W(r,t)={1\over r r_{0}}\sum_{n=0}^{+\infty}e^{-Tx_{n}^{2}t}A_{n}^{2}\sin(x_{n}r)\sin(x_{n}r_{0}),
\label{devl30}
\end{equation}
with
\begin{equation}
A_{n}^{2}={x_{n}\over 2\pi (x_{n}-\sin x_{n}\cos x_{n})}.
\label{devl31}
\end{equation}

If we consider the pure diffusion of a particle in a box, the boundary condition (\ref{devl25}) reduces to $\phi'(1)=0$ and the  $x_{n}$ are solutions of the implicit equation
\begin{equation}
\tan(x_{n})={x_{n}}.
\label{devl32}
\end{equation}
In particular, $x_{0}=0$. This implies that the probability $W(r,t)$
converges for large times to a {\it uniform} profile
$W(r,+\infty)={3/4\pi}$ which is indeed solution of the diffusion
equation in a box. If gravity is taken into account, its first order
effect (in the limit $T\rightarrow +\infty$) is to change the boundary
condition to Eq. (\ref{devl25}). It is {\it as if} we had a diffusion
across the box \cite{chandra,planete} although the true physical
process is a capture by the central mass. The eigenvalues are now
determined by Eq. (\ref{devl28}). The $x_{n>1}$ are hardly modified
(to first order) with respect to the preceding problem but $x_{0}$ is
now different from zero. To first order, we find that $x_{0}^{2}=3/T$
so that $\lambda_{0}=3$ in agreement with the result of
Sec. \ref{sec_largetime}. We also note that $A_{0}^{2}=T/4\pi$ while
$A_{n>0}$ are independent on $T$ (to leading order) and given by
Eqs. (\ref{devl31}) and (\ref{devl32}).

Using Eqs. (\ref{devl5}) and (\ref{devl6}), the probability that the
particle has been captured by the central mass at time $t$ is given by
\begin{equation}
Q(t)={4\pi\over T}\sum_{n=0}^{+\infty}{1-e^{-Tx_{n}^{2}t}\over x_{n}}A_{n}^{2}\overline{\sin(x_{n}r_{0})\over r_{0}}.
\label{devl33}
\end{equation}
If we average over initial conditions with the weight $3r_{0}^{2}$ (uniform distribution), we find  to leading order in $T^{-1}$ that
\begin{equation}
\overline{\sin(x_{n}r_{0})\over r_{0}}=0, \qquad {\rm for }\ n>0, 
\label{devl34a}
\end{equation}
\begin{equation}
\overline{\sin(x_{0}r_{0})\over r_{0}}=x_{0}. 
\label{devl34b}
\end{equation}
Hence, the modes $n>0$ cancel out. Therefore, in the high temperature regime, the probability of capture is given by
\begin{equation}
Q(t)=1-e^{-3t},
\label{devl35}
\end{equation}
for all times.  

The case $D=2$ can be treated by a similar method. Instead of Eqs. (\ref{devl30}) (\ref{devl31}) and (\ref{devl28}), we get
\begin{equation}
W(r,t)=\sum_{n=0}^{+\infty}e^{-Tx_{n}^{2}t}A_{n}^{2} J_{0}(x_{n}r) J_{0}(x_{n}r_{0}),
\label{devl36}
\end{equation}
\begin{equation}
A_{n}^{2}={1\over \pi \lbrack J_{1}^{2}(x_{n})+J_{0}^{2}(x_{n})\rbrack},
\label{devl37}
\end{equation}
\begin{equation}
{x_{n}J_{1}(x_{n})\over J_{0}(x_{n})}={1\over T},
\label{devl38}
\end{equation}
where $J_{n}$ is the Bessel function of order $n$. For the pure diffusion process, $x_{n}=\alpha_{1n}$ are the zeros of $J_{1}$. If gravity is taken into account, then $x_{n>1}\simeq \alpha_{1n}$ while $x_{0}^{2}=2/T$ establishing $\lambda_{0}=2$. The probability that the particle has been captured by the central mass at time $t$ is given by
\begin{equation}
Q(t)={2\pi\over T}\sum_{n=0}^{+\infty}{1-e^{-Tx_{n}^{2}t}\over x_{n}^{2}}A_{n}^{2}\overline{J_{0}(x_{n}r_{0})}.
\label{devl39}
\end{equation}
If we average over the initial conditions with a weight $2r_{0}$ (uniform distribution), we get $\overline{J_{0}(x_{n}r_{0})}=0$ if $n>0$ and $\overline{J_{0}(x_{0}r_{0})}=1$. Therefore, in the high temperature regime, the probability of capture is given, for all times, by
\begin{equation}
Q(t)=1-e^{-2t}.
\label{devl40}
\end{equation}

Finally, for $D=1$, we obtain  
\begin{equation}
W(r,t)=\sum_{n=0}^{+\infty}e^{-Tx_{n}^{2}t}A_{n}^{2} \cos (x_{n}r) \cos (x_{n}r_{0}),
\label{devl41}
\end{equation}
\begin{equation}
A_{n}^{2}={1\over  \lbrack 1+{\sin(2x_{n})\over 2 x_{n}}\rbrack},
\label{devl42}
\end{equation}
\begin{equation}
x_{n}\tan(x_{n})={1\over T}.
\label{devl43}
\end{equation}
For the pure diffusion process, $x_{n}=n\pi$. If gravity is taken into account, then $x_{n>1}\simeq n\pi$ while $x_{0}^{2}=1/T$ establishing $\lambda_{0}=1$. The probability that the particle has been captured by the central mass at time $t$ is given by
\begin{equation}
Q(t)={2\over T}\sum_{n=0}^{+\infty}{1-e^{-Tx_{n}^{2}t}\over x_{n}^{2}}A_{n}^{2}\overline{\cos(x_{n}r_{0})}.
\label{devl44}
\end{equation}
If we average over the initial conditions with a weight $1$ (uniform distribution), we get $\overline{\cos (x_{n}r_{0})}=0$ if $n>0$ and $\overline{\cos (x_{0}r_{0})}=1$. Therefore, in the high temperature regime, the probability of capture is given, for all times, by
\begin{equation}
Q(t)=1-e^{-t}.
\label{devl45}
\end{equation}

\end{document}